\documentclass[english]{article}
\usepackage[T1]{fontenc}
\usepackage[latin9]{inputenc}
\usepackage{float}
\usepackage{graphicx}
\usepackage{esint}

\makeatletter
\newcommand{\lyxaddress}[1]{
	\par {\raggedright #1
	\vspace{1.4em}
	\noindent\par}
}

\usepackage{babel}
\usepackage[superscript,biblabel]{cite}

\makeatother

\usepackage{babel}
\begin{document}
\title{Absence of demixing in 2D binary mixtures of alcohols}
\author{Lydia CHELLI and Aurélien PERERA\thanks{aup@lptmc.jussieu.fr}}
\maketitle

\lyxaddress{Laboratoire de Physique Théorique de la Matière Condensée (UMR CNRS
7600), Sorbonne Université, 4 Place Jussieu, F75252, Paris cedex 05,
France.}
\begin{abstract}
Binary mixtures of two-dimensional, site-based models of alcohols
are investigated by computer simulations, focusing on ideal mixing,
local clustering and miscibility trends. Four systems are considered:
methanol--ethanol, butanol-pentanol, methanol--pentanol, and methanol--octanol,
the first two being expected to be ideal mixtures because of the proximity
of the molecules, while the last two are examples of mixing distant
alcohols. The models retain chemical specificity of their 3D counterparts,
thus allowing one to investigate dimensional constraints and non-trivial
micro-structurations. It is found that mixtures of short and long
alcohols remain miscible, in contrast with the corresponding macroscopic
phase separation observed in three dimensions. This issue is analyzed
through snapshots, site-site distribution functions, structure factors
and Kirkwood--Buff integrals (KBI). It is found that the short range
part features are dominated by charge ordering of the hydroxyl head
groups, while the long range part develops persistent oscillatory
features affecting the determination of the KBI, and suggesting the
presence of anomalous domain-like concentration fluctuations. All
these findings suggest a complex interplay between charge ordering
of the hydrophilic head groups and the hydrophobic tail groups.
\end{abstract}
\textbf{Keywords}: alcohols; mixtures; micro-segregation; two-dimension;
Monte-Carlo simulations; Kirkwood-Buff Integrals.

\section{Introduction}

Since Onsager's celebrated exact solution of the two-dimensional Ising
model \cite{Onsager_Ising,Ising2D_Kaufman}, two-dimensional models
of real three-dimensional systems have played a distinctive role in
statistical physics. In the physics of liquids, two-dimensional systems
have likewise been extensively investigated, both as simplified models
for exploring fundamental questions of liquid-state theory\cite{Alder-Wainwright,Viellard-Baron,Lado-2DHS,DillReview}
and in connection with adsorbed and interfacial fluids and fluids
confined in quasi-two-dimensional geometries\cite{2D-adsorb3,2D-adsorb1,2D-adsorbed2}.
More generally, dimensional reduction provides a useful setting in
which the interplay between molecular interactions, local structure,
and collective fluctuations can be examined under altered geometrical
constraints. Two-dimensional liquids may thus serve as probes for
identifying which features of liquid structure are robust with respect
to dimensionality and which are not.

The purpose of the present paper is two-fold. On one hand it presents
a Monte Carlo simulation study of binary mixtures of two-dimensional
alcohols, as a continuation of our previous study of single components
in Ref.\cite{AUP-2Dalc}. On the other hand, this study focuses on
the concentration fluctuations in these mixtures, particularly since
it is found herein that mixtures of strongly dissimilar alcohols,
such as methanol-octanol for instance, do not demix in two dimensions,
in contrast with real three-dimensional alcohols.

In what concerns the first point, and as in Ref.\cite{AUP-2Dalc},
this paper focuses mostly on the local structure, that is the chain-like
aggregates of the hydroxyl head groups, which were found to be similar
to that of real alcohols. One particular feature in binary mixtures
of alcohols is the ideality they exhibit when they are ``close'',
such as for instance methanol-ethanol, or butanol-pentanol. We previously
studied this property in the case of real alcohols\cite{Pozar2016,AUP-idealMix2}
, and found that, in the case of ideal mixtures, the chain aggregates
are themselves mixtures of both alcohol head groups. In particular,
the small-$k$ pre-peak features of the structure factors are preserved
upon mixing conditions. This part of the study also serves to showcase
the two dimensional site-interaction models, which has been previously
tested in the case of water\cite{SS-water,2D_MBwater_Dill} and neat
alcohols\cite{AUP-2Dalc}. These models can be compared with the original
two dimensional models which are based on the Mercedes-Benz interaction
model,\cite{BenNaim-MB,2D_MBwater_Dill} and their derivatives\cite{Tomaz-roseWater},
which are purely orientational models. Real 3D molecular models, such
as the OPLS\cite{Jorgensen1996} or TraPPE\cite{trappe_Siepmann}
force fields are based on atomic site-site interactions. It is therefore
advantageous to develop site-based interaction models for 2D molecular
liquids, instead of the MB class models.

In what concerns the second point, snapshots of all mixtures show
features suggesting micro-segregation, but this is not always reflected
by a distinct domain pre-peak in the structure factors, which may
instead display a rise towards $k=0$.

These two issues are connected through the labile chain formation
of the hydroxyl groups and its relation to concentration fluctuations
and aggregate formation. In particular, the same site-site correlation
functions that reveal local structure also display anomalous long-range
behaviour associated with micro-heterogeneity. This connects with
the duality between concentration fluctuations and micro-heterogeneity
explored in our previous work and summarized in Section\ref{subsec:The-conundrom-of}.

The structure of the paper is the following. In the next section we
summarise the force field and computational details, together with
a recap of the issues with fluctuations in micro-segregated mixtures.
This latter section showcases this investigation in the midst of our
previous ones. The result section showcase the structure of these
mixtures through snapshots, pair distribution functions and related
structure factors. The problem with the running Kirkwood-Buff integrals
are discussed in Section\ref{sec:Domain-order-characteristics} .
Finally, a discussion section is followed by our conclusions.

\section{Model, theoretical and computational details}

\subsection{2D alcohol models}

\begin{figure}[H]
\centering
\includegraphics[scale=0.3]{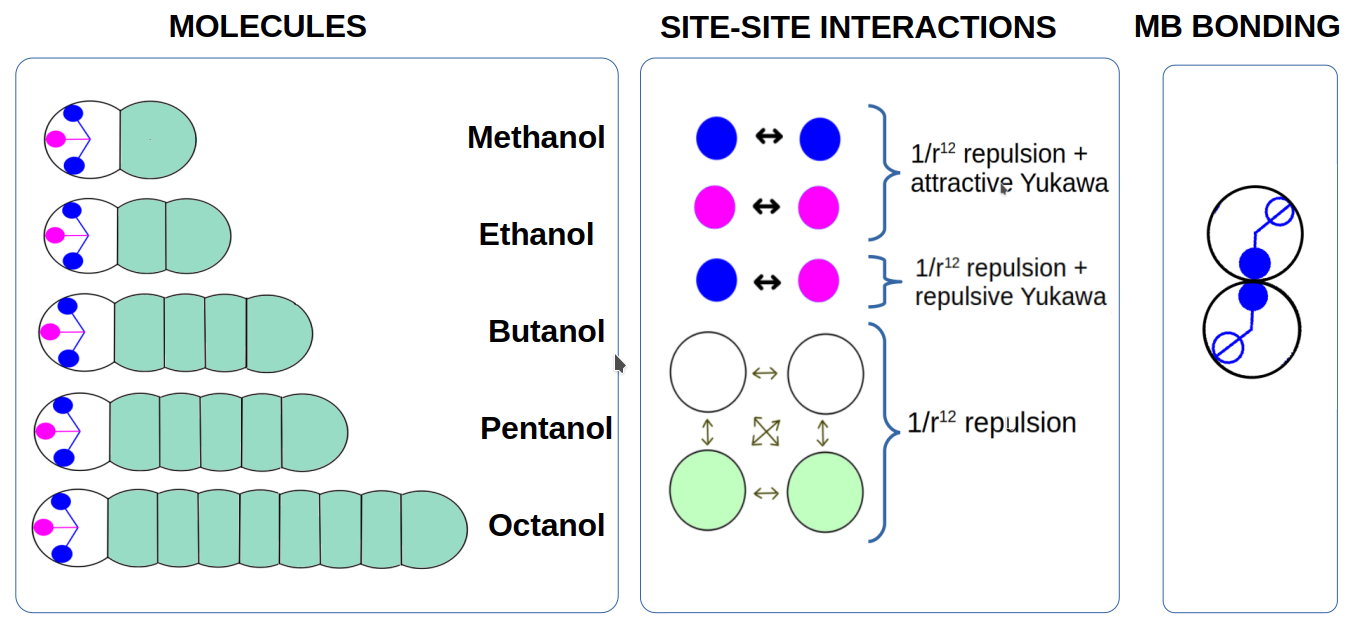}

\caption{Illustration of the 2D alcohol models used in this work (left panel),
the site-site interactions (middle panel) and Mercedes-Benz (MB) type
hydrogen bonded dimer for 2D alcohols (right panel). See Tables S1
and S2 in the SI document.}
\label{FigModels}
\end{figure}

The interaction-site-based two dimensional alcohol models follow the
same patterns as previously developed for the 2D site-site SSMB model
of water\cite{SS-water}. Namely, we replace the interaction between
the MB arms by ``charged'' site interactions, combined with a repulsive
interaction site. This is illustrated in Fig.\ref{FigModels}

The hydroxyl head group is modeled as a 3 arms state, with 2 attractive
sites (in blue in Fig.\ref{FigModels}), which represent both the
oxygen and hydrogen sites. We have followed Ref.\cite{AUP-2Dalc}
and made the two sites of equal valence. This is also consistent with
the MB symmetry of the 2D water model, for which the oxygen and hydrogen
sites are not distinguished and simply replaced by similar attractive
sites. This is fully in the spirit of the original MD model \cite{BenNaim-MB,DillReview},
where the 3 arms are identical. This seemingly nonphysical representation
of real water is justified by the fact that such a simple model is
able to capture the physical properties of the associating liquids
as well as some of the anomalies of real water, such as the density
maximum, the compressibility minimum and the two types of amorphous
phases (dense HDL and expanded LDL)\cite{DillReview}. In order to
avoid misalignments of the 2 arms, it is necessary to introduce a
single repulsive site (in magenta in Fig.\ref{FigModels}). The interactions
are chosen similarly to our previous work for water, with a $1/r^{12}$
dispersion repulsion and a Yukawa form for hydrogen-bonding part of
the site-site potential:
\begin{equation}
v_{ij}(r)=4\epsilon_{ij}\left(\frac{\sigma_{ij}}{r}\right)^{12}+\alpha_{ij}\frac{Z_{i}Z_{j}}{r}\exp\left[-\frac{(r-R_{ij})^{2}}{\kappa_{ij}}\right]\label{V12}
\end{equation}
where we use the Lorentz rule for the site diameters $\sigma_{ij}=(\sigma_{i}+\sigma_{j})/2$
and the Berthelot rule for the energy parameters $\epsilon_{ij}=\sqrt{\epsilon_{i}\epsilon_{j}}$,
where $\sigma_{i}$ and $\epsilon_{i}$ are the diameter and energy
of site $i$, respectively. For the screened Coulomb-Yukawa part $Z_{i}$
is the valence of site i, and $\alpha_{ij}$ is the sign of the interaction
between sites $i$ and $j$. The alkyl tail atoms are modeled as zero
valence sites $Z_{A}=0$. 

The parameters of the interactions are displayed in Tables S1 and
S2 of the Supplemental Information (SI) document.

\subsection{Computational details\protect\label{subsec:Computational-details}}

As in Ref.\cite{AUP-2Dalc} for the neat alcohols, NVT Monte Carlo
simulations have been performed, with a total number of molecules
around $N=1000$, depending on the initial packing conditions. The
packing fraction of all the systems is kept constant at $\eta=0.6,$which
corresponds to a dense liquid regime. The temperature is fixed at
$T=2$, which corresponds to a mid range liquid state temperature
for the neat alcohols studied in Ref.\cite{AUP-2Dalc}. Other temperatures
have been studied, as discussed below in the results sections. Typical
solute mole fractions $x=0.2$, $x=0.5$ and $x=0.8$ have been studied,
corresponding to a range between solvent-rich and solute-rich regimes.
Each system was melted at the high temperature $T=15$ for about $100$
thousand steps, before brought in successive steps to $T=2$. Then
equilibration runs of 100 thousand Monte Carlo steps have been performed,
each such step consisting of trial moves of all the N molecules. Several
production runs of 150 thousand steps were performed. It was necessary
to do so in order to examine whether the long range domain oscillations
in the tail of the site-site pair correlation functions $g_{a_{i}b_{j}}(r)$
satisfy self-averaging, as discussed in the results sections below.
This is a central point of the present study and is easier to study
in 2D model systems than in 3D realistic systems. We reported similar
problems in several previous works Refs.\cite{Perera2017a,Perera2022,myIUPAC},
although their statistical nature was difficult to assess properly. 

In addition, a problem noticed in our previous study of pure alcohol
is enhanced in the present case of mixtures: it concerns the ringing
effect in the low $k$ part of the structure factors. This ringing
artifact arises from the combined effects of the statistical noise
in the tails of the $g_{a_{i}b_{j}}(r)$, the requirement to interpolate
these functions onto a logarithmic gird before performing the 2D Fourier-Talman
transform, and the long range domain oscillations of the tails. Their
smaller magnitude in the case of neat alcohol is explained by the
absence of the domain oscillations. Thus, the amplified ringing effects
are essentially coming from the domain oscillations, which themselves
are reminiscent of non-self-averaging. This problem implies that domain
correlations requires further numerical investigations. For these
reasons, we have made no attempts to correct them in the present study. 

\subsection{The conundrum of ``structure formation'' \protect\label{subsec:The-conundrom-of}.}

In a homogeneous liquid system the microscopic 1-body density of species
$a$ with mean density $\bar{\rho}_{a}$ can be written as
\[
\rho_{a}(1)=\sum_{i_{a}}\delta(1-i_{a})=\bar{\rho}_{a}+\delta\rho_{a}(1)
\]
where we adopt the convention $1=(\mathbf{r}_{1},\Omega_{1})$ and
the index $i_{a}$ designates both the particle $i$ of species $a$
and its generalized coordinates $i_{a}=(\mathbf{r}_{i_{a}},\Omega_{i_{a}})$.
The key point of homogeneity here is that, at the one-body level one
has\cite{HansMac,chaikin2000principles}
\[
<\delta\rho_{a}(1)>=0
\]
where <.> designates an ensemble average, thereby removing any trace
of local heterogeneity of species $a$. However, at pair level, one
has\cite{HansMac,chaikin2000principles}
\[
<\delta\rho_{a}(1)\delta\rho_{a}(2)>=\bar{\rho}_{a}^{2}\left[g_{aa}(1,2)-1\right]
\]
thus transferring any signature of local heterogeneity of species
$a$ to the pair-correlation level.

If there is sustained heterogeneity, such as species a forming specific
domains, then it is likely that the corresponding domain correlations
would show up as oscillatory features in the long-range part of the
angle-averaged correlation function $g_{aa}(r)$, and hence translated
into a domain pre-peak in the corresponding angle-averaged structure
factor $S_{aa}(k)$.

Problems arise when the micro-segregation does not leave a ``proper''
oscillatory trace in the long range correlations of $g_{aa}(r)$.
In this case, micro-segregation can be interpreted as concentration
fluctuations, and according to the Kirkwood-Buff theory\cite{Kirkwood1951,Nishikawa2021},
would give rise to a $k=0$ peak (or rise) in the structure factor
$S_{aa}(k)$. This does not necessarily mean that the system is close
to a spinodal or subject to pre-critical fluctuations. This situation
is known to occur in many aqueous mixtures and is often accompanied
by slow relaxation of the heterogeneous domains, and poses a real
conundrum in computer simulations\cite{MyAce1,Overduin2019}.

As will be seen in Section\ref{sec:Domain-order-characteristics},
the binary mixtures studied herein pose exactly this problem of poorly
defined micro-domain structures.

\section{Micro-structure through snapshots}

Snapshots are more useful in 2D than in 3D, because of the direct
surface view without hidden molecular formations. They provide a quick
insight of the preferred structures, especially for these alcohols
which tend to present chain-like aggregation of the hydroxyl groups.
To facilitate the view, the oxygen sites of the first alcohol component
are colored in red and that of the second alcohol (the solute with
mole fraction $x$) are colored in blue. The blue and magenta bonding
sites in Fig.\ref{FigModels} are considered as hydrogens and colored
in white in the snapshots and the alkyl tails are shown as green sticks,
thus narrowing the focus on the hydroxyl aggregates. In addition,
we provide a second set of snapshots (right panels) that distinguishes
species instead of atoms. In this way, one can appreciate the extent
of the micro-separation between the two species.

Below, we first address the ideality of the mixtures, specifically
when the two alcohols are close to each other in terms of alkyl tail
lengths (methanol-ethanol and butanol-pentanol). We have previously
examined this topic for real alcohols in 3D\cite{AUP-idealMix-JCP,AUP-idealMix2},
and the main conclusion was that the enormous difference in strength
between the charge interaction and the dispersion interaction explained
why the chain clusters were clusters of randomly mixed methanol and
ethanol hydroxyl heads, hence acted as an ideal mixture, as evidenced
by near identical oxygen-oxygen pair correlation functions at all
concentrations.

Then, we address the unexpected issue of the absence of demixing in
2D, specifically for methanol-pentanol and methanol-octanol mixtures.
This is not simply a consequence of fluctuations being larger in 2D,
because it is very easy to build a model with molecules A and B interacting
through cross repulsion, such as the Widom-Rowlinson model \cite{WidomRowlinson,WidomRowlinson-simu}.
Any such mixture model will show rapid demixion in simulations. Rather,
it is the consequence of a competition between the energy favored
association of the head-groups and the entropic dispersion of the
uncharged alkyl tails, which is more constrained in 2D than in 3D. 

\subsection{Search for ideality of mixing: methanol-ethanol}

\begin{figure}[H]
\centering
\includegraphics[scale=0.3]{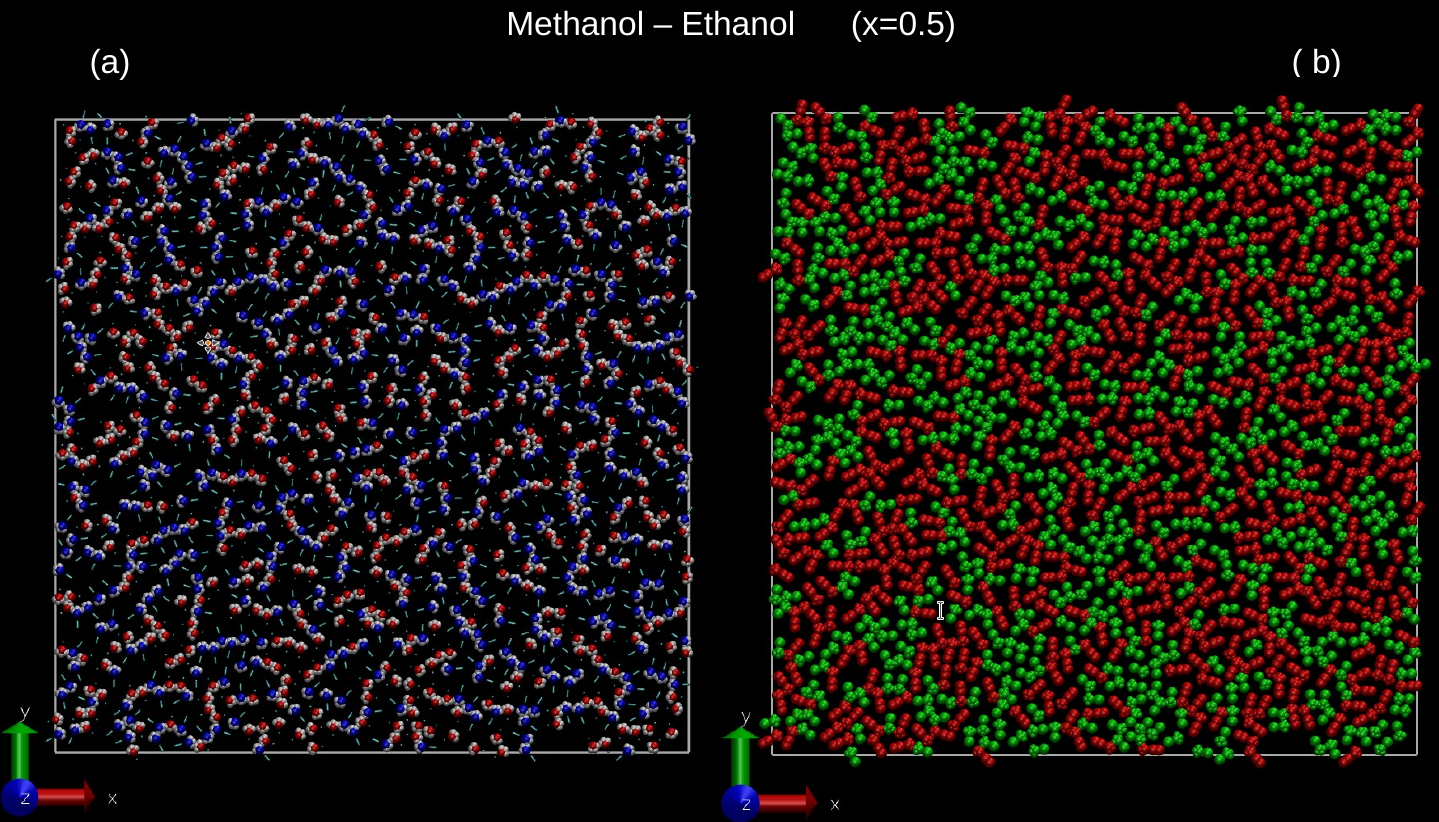}

\caption{Snapshots of the equimolar (ethanol mole fraction $x=0.5$) methanol-ethanol
mixture. Left panel (a) shown the methanol oxygen site in red, the
ethanol oxygen site in blue and the hydrogen(see text) in white. The
alkyl tails are shown as a thin green line in order to better highlight
the chains formed by the head groups. Right panel (b) shows the same
snapshot but with methanol highlighted in green and ethanol in red,
in order to demonstrate the micro-segregation of both species.}

\label{snapMethEth}
\end{figure}

Fig.\ref{snapMethEth} shows in the left panel the snapshots for the
equimolar $x=0.5$ methanol-ethanol mixture, highlighting the hydroxyl
group chain-like clusters and their rich variety (linear, branched).
Similar snapshots are shown in the Supplementary Information (SI)
document for the ethanol mole fraction $x=0.2$ (Fig.SI-1a) and $x=0.8$
(Fig.SI-1b), where it can also be visually seen that the proportion
of mixing of the red and blue sites follows the concentrations of
each species. 

The right panel shows the same snapshot, but with methanol shown in
green and ethanol in red. This way we can observe the micro-segregation
between the two species, with distinct patches of each species. But
it could also be interpreted as large concentration fluctuations.

The mixing within the chain-like aggregates, which is visible for
$x=0.5$ enforces the idea of ideal mixing. For other concentrations,
the mixing appears to follow the concentration pattern, further supporting
the hypothesis of ideal mixing.

\subsection{Search for ideality of mixing: butanol-pentanol}

\begin{figure}[H]
\centering
\includegraphics[scale=0.3]{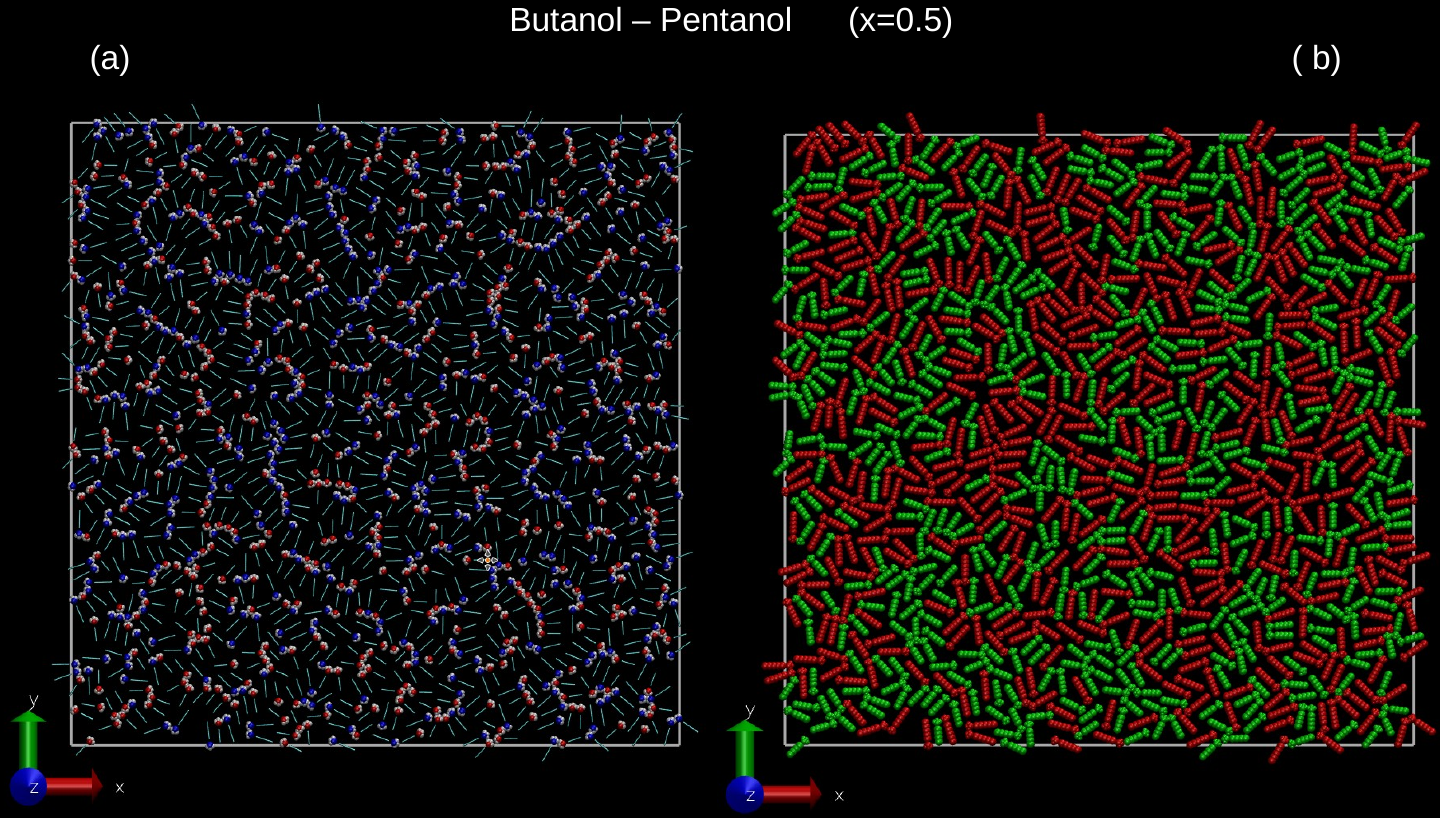}

\caption{Snapshot of the equimolar butanol-pentanol mixtures with the same
conventions as in Fig.\ref{snapMethEth}}

\label{snapButPent}
\end{figure}

Fig.\ref{snapButPent} shows the snapshots for various butanol-pentanol
mixtures, with the left panel showing chain clusters somewhat shorter
than for methanol-ethanol. This trend can directly be associated with
the hindering effect of the longer alkyl tails.

It is somewhat visually apparent that the mixing of the blue and red
atoms within each chain is not similar to that of methanol-ethanol
mixtures, since one observes many clusters with nearly single color
head groups. Other concentrations are shown in Fig.SI-2a and Fig.SI-2b.
Similar trends are observed for other close mixtures, such as hexanol-heptanol
or heptanol-octanol, for instance. This means that ideal mixing between
close 2D alcohols is maintained when the alkyl tails are made longer,
although it becomes less apparent at the local chain level.

The second snapshot again shows a clear micro-segregation of both
species, where the longer pentanol (red) appears to form better-defined
packs than butanol (green). This local heterogeneity explains why
mixing of the hydroxyl head groups between the two species becomes
difficult.

\subsection{Mixing instead of demixing: methanol-pentanol}

\begin{figure}[H]
\centering
\includegraphics[scale=0.3]{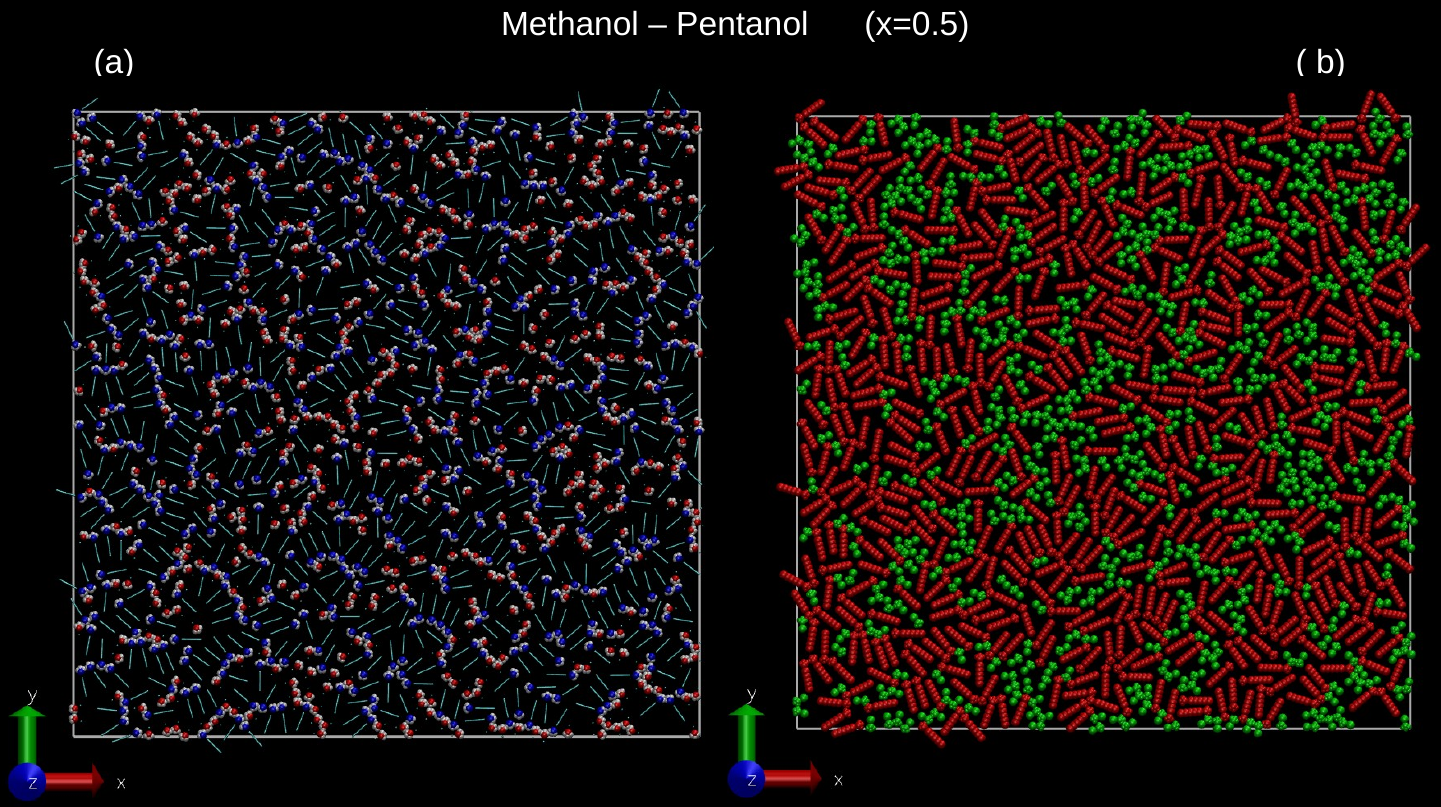}

\caption{Snapshot of the equimolar methanol-pentanol mixtures with the same
conventions as in Fig.\ref{snapMethEth}}

\label{snapMethPent}
\end{figure}

We now examine mixtures of alcohols are that farther apart in terms
of the alkyl-tail length. For real alcohols, methanol is not fully
miscible with pentanol and longer alcohols. Conversely, in 2D, Fig.\ref{snapMethPent}
shows no phase separation for pentanol concentration $x=0.5$. Similar
results are reported in the SI document for $x=0.2$ and $x=0.8$,
indicating that 2D methanol and pentanol remain fully miscible. Instead,
one observes chain-like hydroxyl head group clusters with topology
more or less similar to that observed when mixing similar alcohols.
For real 3D alcohol mixtures, the smaller alcohols can cluster more
efficiently because the short tails have more room to rotate, thus
leaving the longer alcohols phase separated. In 2D, this rotation
of the short tails is constrained to the plane, hence forcing mixing.
In addition, it is energetically more efficient to cluster the hydroxyl
heads together, favoring the mixing along the chains. 

This is a striking example of mixing favored not in the bulk but within
chain clusters. The alkyl tails act as entropic constraints leading
to the observed rich topology. Similar figures for pentanol concentrations
$x=0.2$ and $x=0.8$ are shown in the SI document in Fig.SI-3a and
Fig.SI-3b, respectively.

The second figure on the right again shows typical micro-segregation
pattern.

\subsection{Mixing instead of demixing: methanol-octanol}

\begin{figure}[H]
\centering
\includegraphics[scale=0.3]{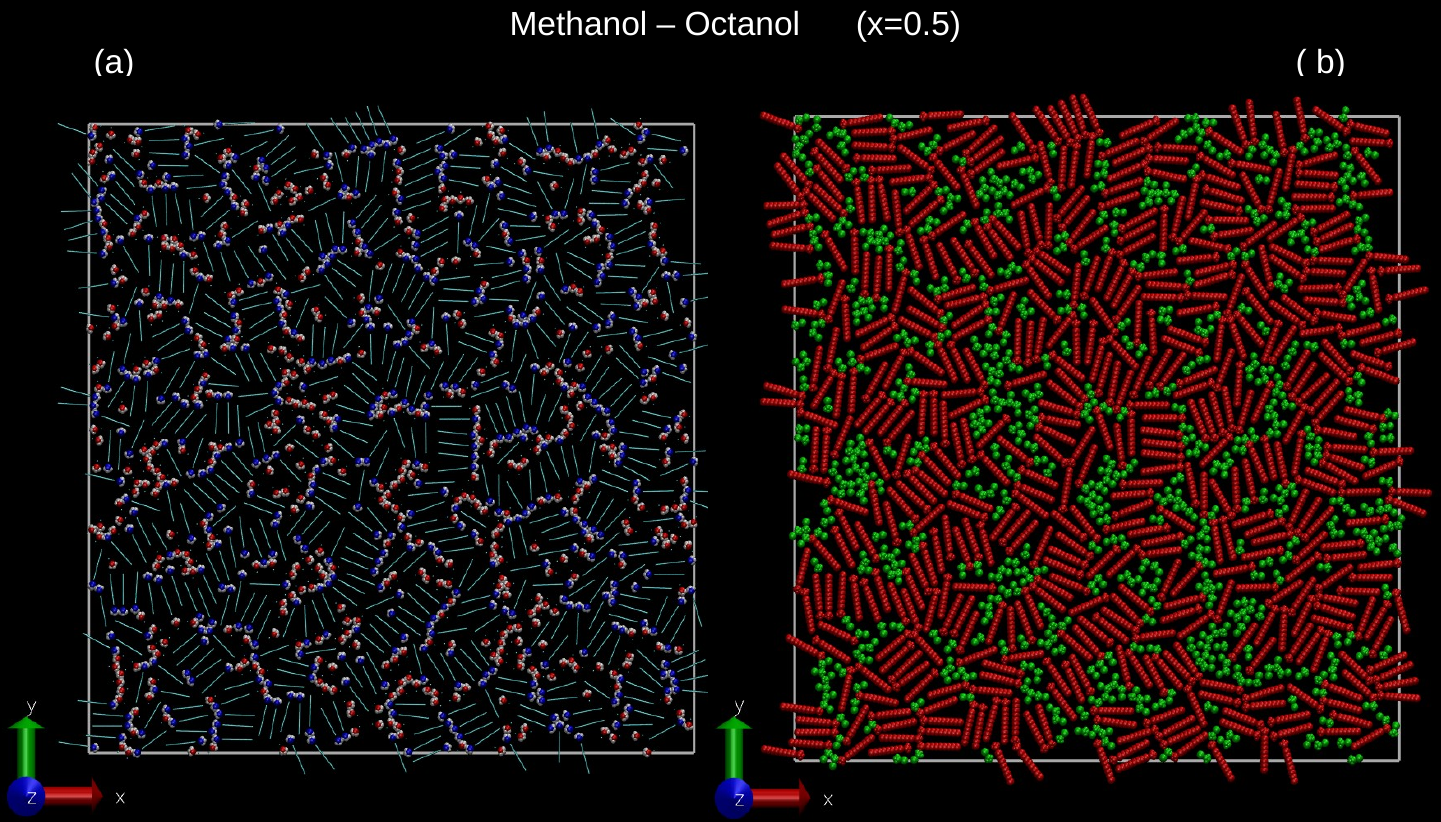}

\caption{Snapshot of the equimolar methanol-octanol mixtures with the same
conventions as in Fig.\ref{snapMethEth}}

\label{snapMethOct}
\end{figure}

The form of mixing within chain clusters leads to a somewhat more
complex form of micro-structure in methanol-octanol mixtures, when
one can observe now the hydroxyl head-groups chain-like domains segregating
from the tail domains, most visible for octanol mole fraction $x=0.5$
in the right panel of Fig.\ref{snapMethOct}.

Overall, mixing inside chains remains more or less similar to that
observed in previous cases. Similar figures for octanol concentrations
$x=0.2$ and $x=0.8$ are shown in the SI document in Fig.SI-4a and
Fig.SI-4b, respectively.

\section{Structural properties}

What trends are expected in the $g(r)$ and $S(k)$ functions? We
distinguish three main features: ideality, chain-like aggregation,
and micro-heterogeneity. In ideal mixtures of real 3D alcohols \cite{AUP-idealMix-JCP,AUP-idealMix2,Lovrincevic2019},
the oxygen-oxygen distribution functions are nearly superimposed and
concentration independent, reflecting the statistical mixing of both
species within the hydroxyl chains. Chain-like aggregation, in turn,
produces the characteristic depletion of the oxygen-oxygen correlations
beyond the first neighbour shell, as previously observed in both 2D
and 3D neat alcohols. Finally, micro-heterogeneous domains may produce
small-amplitude long-range oscillations associated with domain-domain
correlations. These three signatures will guide the structural analysis
below, in connection with the snapshots of the previous section. We
do not report data for methanol-pentanol since it is quite similar
to methanol-octanol.

In addition, the SI document provides charge order correlation functions
between the ``charge'' bearing blue and pink sites shown in Fig.\ref{FigModels},
and many model mixtures and typical concentrations, in Fig.SI-5-7.
End atom correlations are equally provided in Fig.SI-8-9 in order
to appreciate the role of the hydrophobic alkyl tails.

\subsection{Methanol-ethanol}

\begin{figure}[H]
\centering
\includegraphics[scale=0.3]{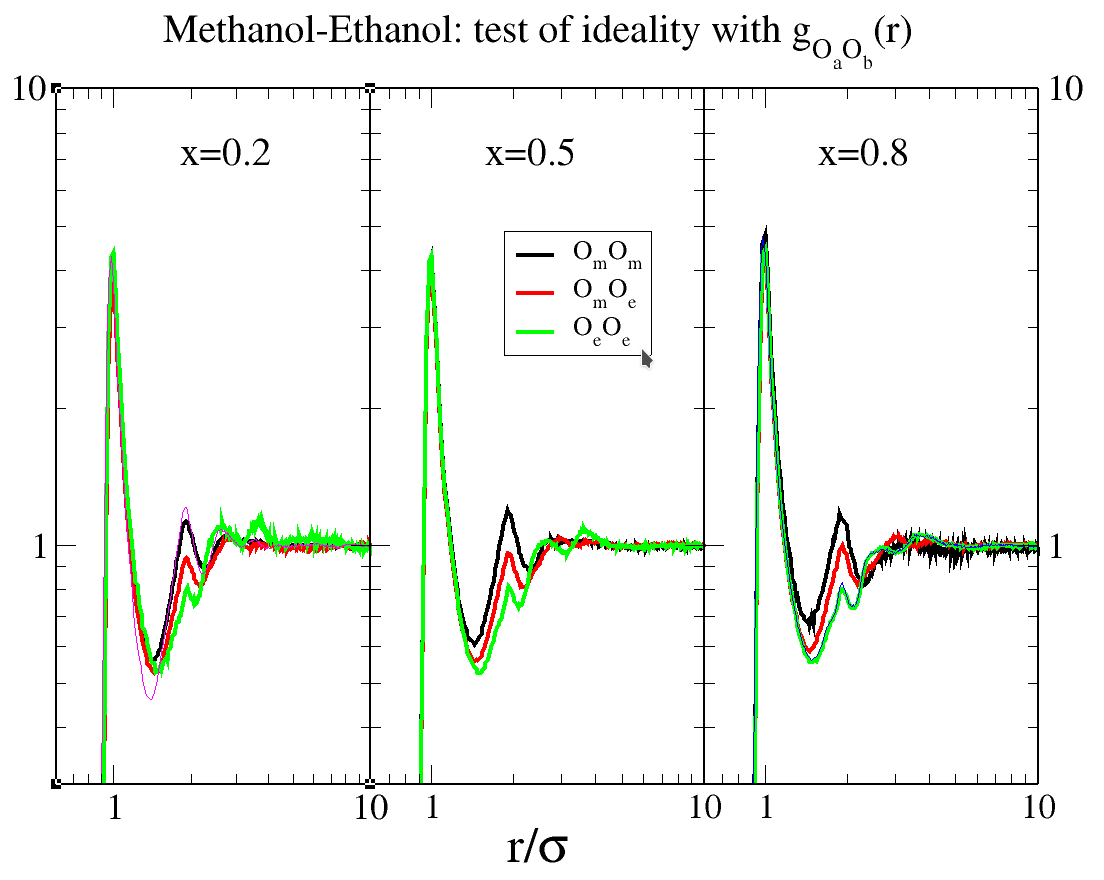}

\caption{Log-log plot of the oxygen-oxygen pair correlation functions $g_{O_{a}O_{b}}(r)$
in methanol-ethanol mixtures and for typical ethanol concentrations
$x=0.2$ (left panel), $x=0.5$ (middle) and $x=0.8$ (right). The
methanol oxygen correlations $g_{O_{m}O_{m}}(r)$ are shown in black
lines, ethanol oxygen correlations $g_{O_{e}O_{e}}(r)$ in green,
and cross correlations $g_{O_{m}O_{e}}(r)$ in red. The correlation
for pure methanol and ethanol are shown in thin magenta and blue curves
in the left panel and in the right panel, respectively.}

\label{GR-MethEth}
\end{figure}

Fig.\ref{GR-MethEth} compares the oxygen-oxygen pair correlation
functions for 3 different ethanol mole fractions. The various atom-atom
$g_{O_{a}O_{b}}(r)$ are shown in log-log plot, in order to keep the
large differences between peak structures at a similar range for all
distances. This tends also to magnify the differences at short range.
The neat methanol and ethanol $g_{O_{a}O_{a}}(r)$ are equally shown
in thin lines in the first and last panels. As expected from the snapshots
of Fig.\ref{snapMethEth}, we clearly observe two of the three features
described above, indicating near perfect ideal mixing. The first peaks
are of nearly equal heights and superimposable, denoting independence
with respect to the whether the oxygen atoms belong to methanol or
ethanol. The correlation of second neighbour and high ones are depleted,
consistent with chain formation of the hydroxyl heads groups. There
is some difference here, since methanol and ethanol do not have the
same type of depletion, as can be seen from the thin curves. However,
from these plots, we do not see clear domain correlation in the long
range, because these are of much smaller amplitude. If they exist,
these are expected to be seen in the analysis of the structure factor
or the running Kirkwood-Buff integrals in the following sections.

\begin{figure}[H]
\centering
\includegraphics[scale=0.3]{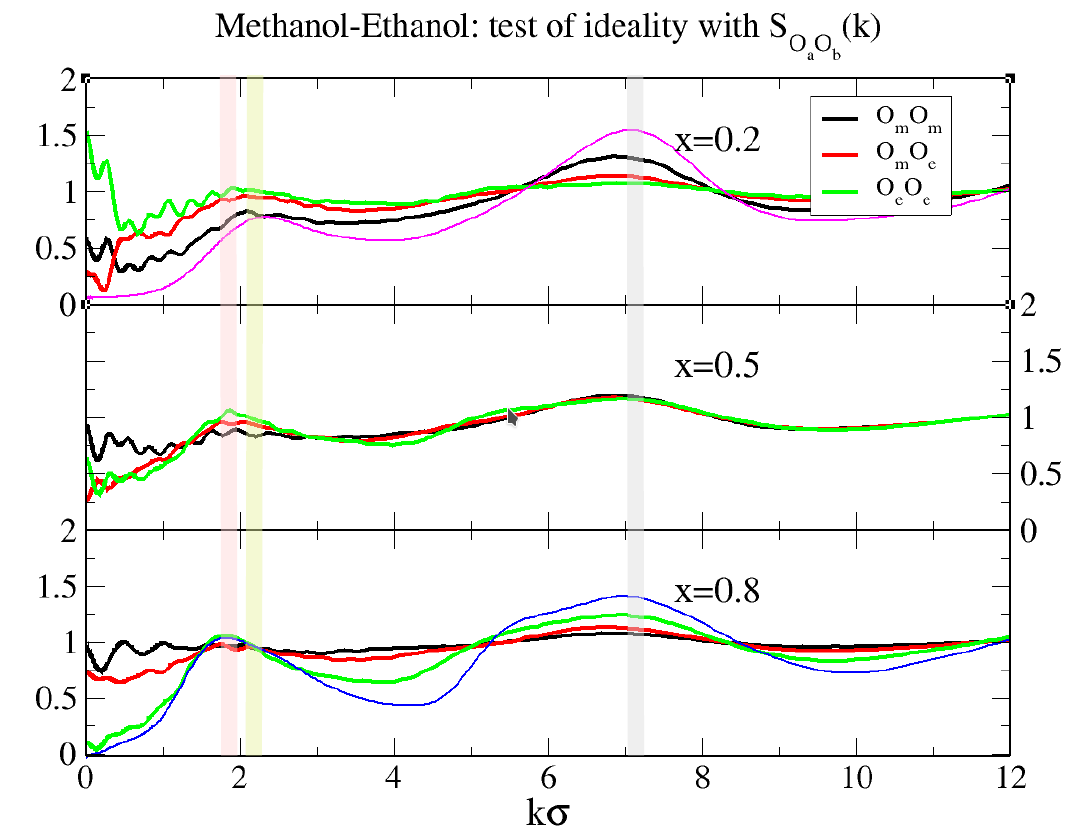}

\caption{Oxygen-oxygen structure factors $S_{O_{a}O_{b}}(k)$ corresponding
to the pair $g_{Oa}O_{b}$(r) shown in Fig.\ref{GR-MethEth}, with
same line and color patterns. The pure methanol and ethanol structure
factors are shown in magenta (upper panel) and blue (lower panel).
The vertical lines indicates the various peaks (see text): pale gray
for the main peak, pale green and red for the pre-peaks of methanol
and ethanol, respectively.}

\label{Sk-MethEth}
\end{figure}

The corresponding oxygen-oxygen structure factors are shown in Fig.
\ref{Sk-MethEth}, together with those of the pure components. The
main peaks at $k\sigma\simeq7$, corresponding to the molecular packing
scale $\ensuremath{\sigma\simeq3}\mathring{A}$, lie near $\ensuremath{k\simeq2}\mathring{A}\ensuremath{^{-1}}$,
as also observed for real alcohols and their mixtures\cite{Pozar2024,my_monool}
.

The pre-peaks at $k\sigma\simeq2.1$ and $1.8$ similarly match those
of pure 2D methanol and ethanol, respectively\cite{AUP-2Dalc} . No
additional domain pre-peak is observed below $\ensuremath{k\simeq2}\mathring{A}\ensuremath{^{-1}}$.
Instead, the structure factors rise towards $k=0$, indicating concentration
fluctuations despite the apparent micro-segregation seen in the snapshots.

The spurious ringing at small $k$, already discussed in Section \ref{subsec:Computational-details}.,
originates from the amplification by the 2D Talman-Fourier transform
of weak irregular oscillations in the long-range tails of $g(r)$.
Their statistical behaviour will be examined in Section \ref{sec:Domain-order-characteristics}
.

The charge-order correlations shown in Figs. SI-5a--SI-5c display
nearly identical first two peaks and almost no long-range correlation,
indicating that they mainly probe the internal coherence of the hydroxyl-chain
aggregates. The end-carbon correlations (Fig. SI-8) are weaker and
shorter-ranged, as expected for sites essentially slaved to the head-group
structure.

\subsection{Butanol-pentanol}

\begin{figure}[H]
\centering
\includegraphics[scale=0.3]{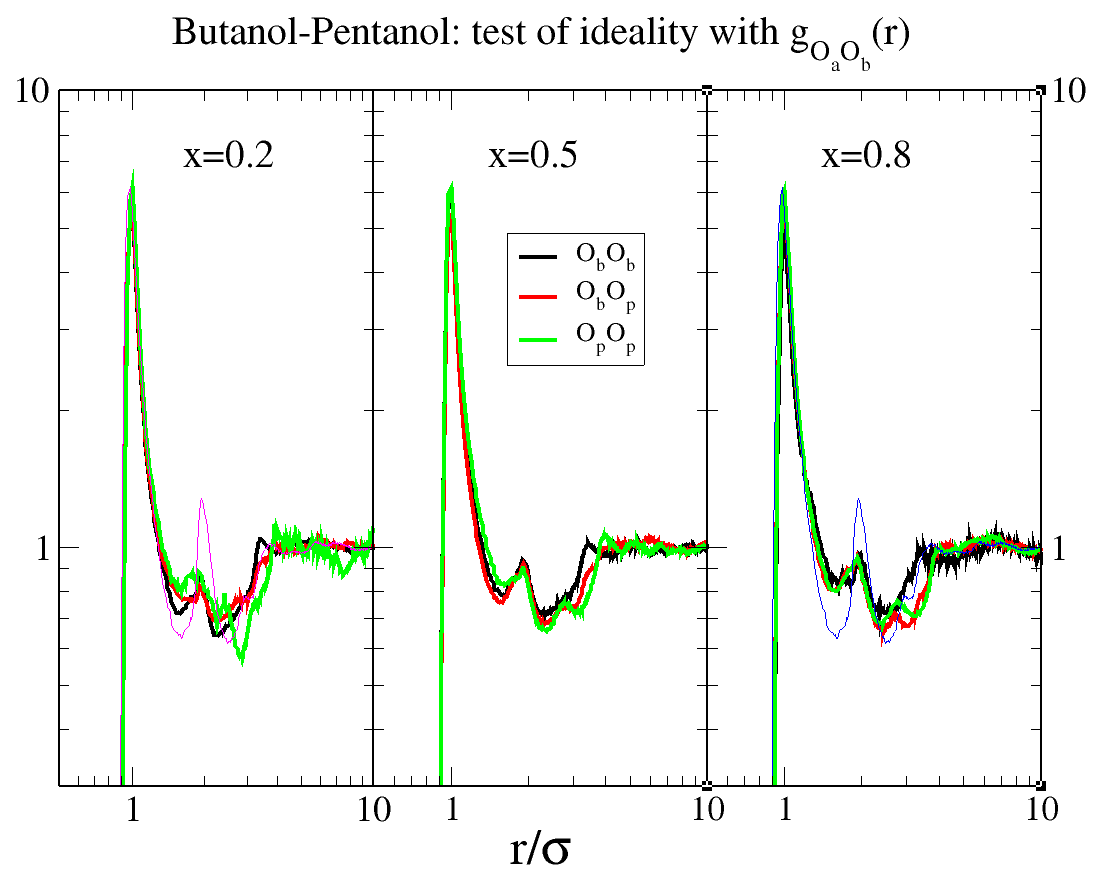}

\caption{Log-log plot of the oxygen-oxygen pair correlation functions $g_{O_{a}O_{b}}(r)$
in butanol-pentanol mixtures, for pentanol mole fractions $x=0.2$,
$0.5$ and $0.8$. Plotting conventions as in Fig.\ref{Sk-MethEth}.}

\label{GR-ButPent}
\end{figure}

The butanol-pentanol mixtures show clear tendencies towards ideal
mixing, as seen from the similarity of the $g_{O_{a}O_{b}}(r)$ in
Fig. \ref{GR-ButPent}, despite the differences magnified by the log-log
representation. 

\begin{figure}[H]
\centering
\includegraphics[scale=0.3]{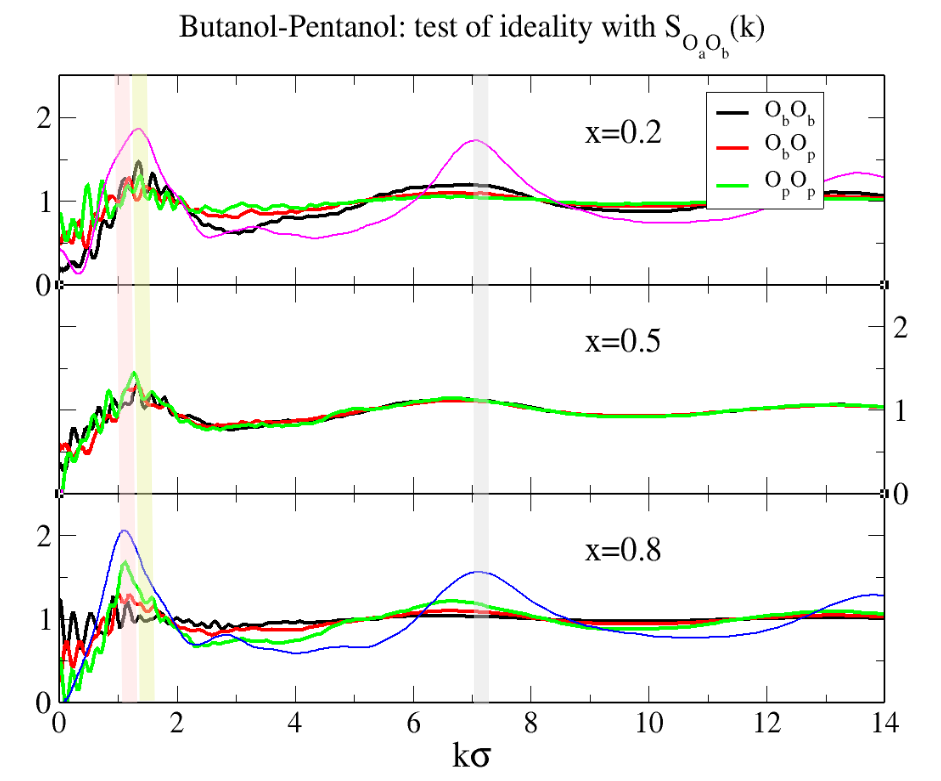}

\caption{Oxygen-oxygen structure factors $S_{O_{a}O_{b}}(k)$ corresponding
to the pair $g_{Oa}O_{b}$(r) shown in Fig.\ref{GR-ButPent}, with
same line and color patterns, but adapted for the butanol-pentanol
mixtures, with pentanol mole fraction $x=0.2$, $0.5$ and $0.8$}

\label{SK-ButPent}
\end{figure}

This is even clearer in the corresponding structure factors (Fig.
\ref{SK-ButPent}), particularly for $x=0.5$. The main peaks and
pre-peaks match those of the neat liquids, while no separate domain
pre-peak is observed, consistently with quasi-ideal mixing of the
polar heads within the chains.

The charge order correlation functions are shown in Fig.SI-5a-c of
the SI document. It can be seen that the quasi ideality of this mixture
is reflected in the close similarity of all the curves, at least closer
that for methanol-ethanol in Figs.SI-6a-c.

More intriguing is the absence of both a domain pre-peak and a $k=0$
rise. In view of the snapshots in Fig.\ref{snapButPent}, we hypothesize
that the characteristic domain size may coincide with the mean hydroxyl-chain
size. This interpretation is supported by the cross-correlation pre-peak
$S_{O_{b}O_{p}}(k)$, which coincides with those of pure butanol and
pentanol.

The last carbon correlation functions are shown in Fig.SI-8. It can
be seen that the 50\% and 80\% cases show a good similarity, hence
we can conclude that ideal mixing holds better when the larger alcohol
concentration dominates. This is of course true only for alcohols
that are are close -eg- hexanol-heptanol for instance. This imposed
ideality appears as an entropic consequence.

\subsection{Methanol-octanol}

\begin{figure}[H]
\centering
\includegraphics[scale=0.3]{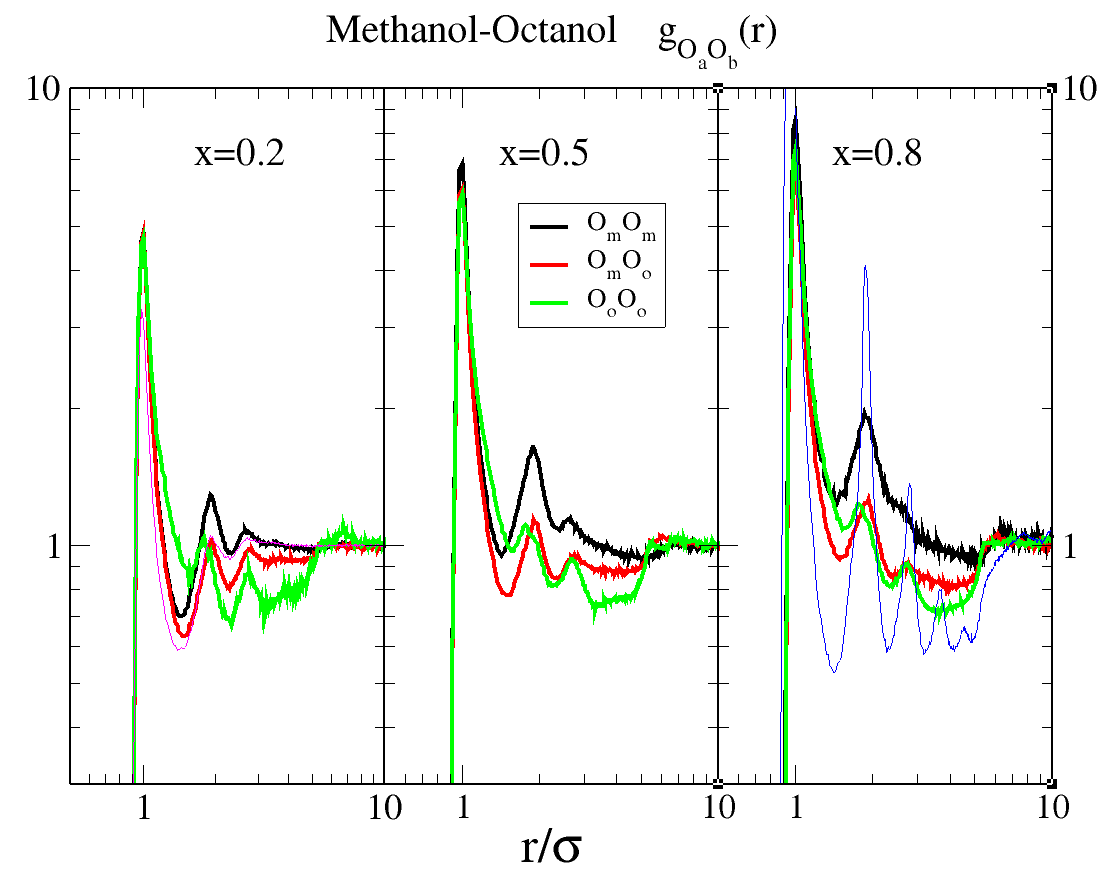}

\caption{Log-log plot of the oxygen-oxygen pair correlation functions $g_{O_{a}O_{b}}(r)$
in methanol-octanol mixtures, for octanol mole fractions $x=0.2$,
$0.5$ and $0.8$. Plotting conventions as in Fig.\ref{Sk-MethEth}.}

\label{GR-MethOct}
\end{figure}

Another interesting case is that of methanol-octanol mixtures, which
remain mixed in 2D, unlike their real 3D counterparts. The oxygen-oxygen
pair correlation functions in Fig. \ref{GR-MethOct} show strong deviations
from ideal mixing. Methanol nevertheless retains the characteristic
depletion beyond the first-neighbour shell, indicating chain formation,
consistently with the snapshots.

\begin{figure}[H]
\centering
\includegraphics[scale=0.3]{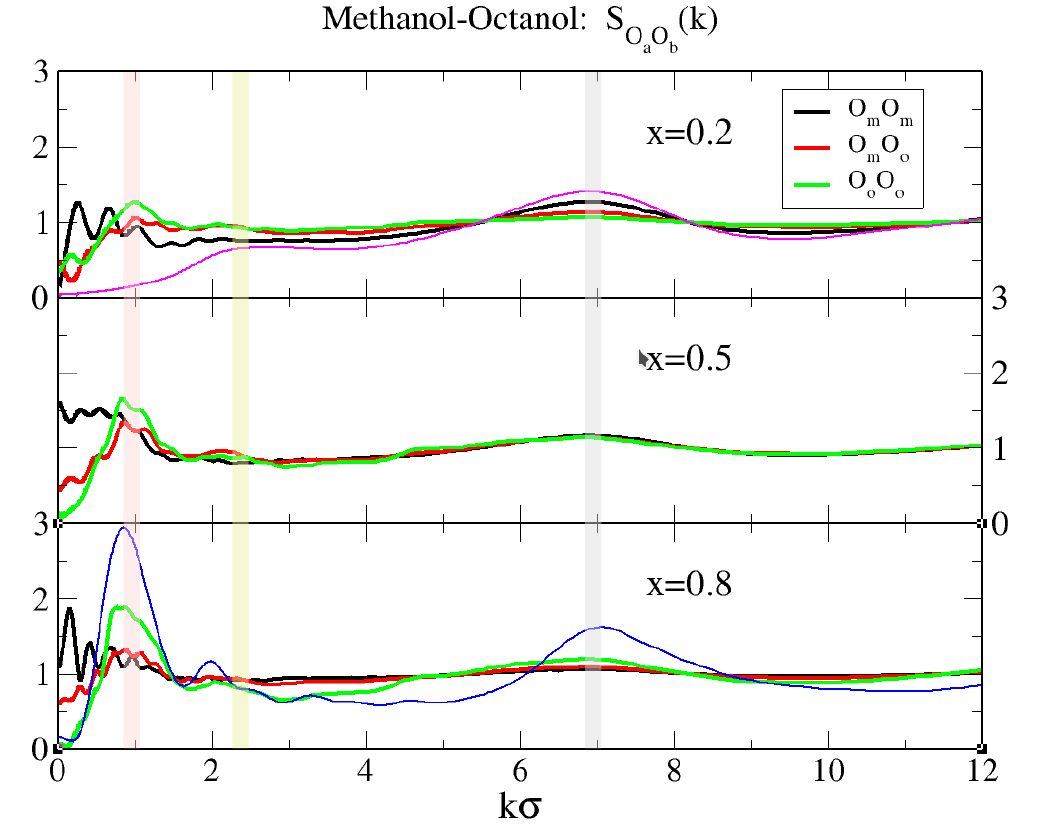}

\caption{Oxygen-oxygen structure factors $S_{O_{a}O_{b}}(k)$ corresponding
to the pair $g_{Oa}O_{b}$(r) shown in Fig.\ref{GR-MethOct}, with
same line and color patterns, but adapted for the methanol-octanol
mixtures, with octanol mole fraction $x=0.2$, $0.5$ and $0.8$.}

\label{SK-MethOct}
\end{figure}

The corresponding structure factors (Fig. \ref{SK-MethOct}) display
the main peaks and respective pre-peaks of both species, but again
no separate domain pre-peak. Only the methanol-methanol structure
factor shows a $k=0$ rise at all concentrations. Interestingly, for
$x=0.5$, a clear feature appears at the octanol pre-peak position
(pink vertical line), suggesting that the characteristic octanol domain
size may coincide with that of the octanol hydroxyl chains. 

The snapshot in Fig. \ref{snapMethOct} supports this interpretation.
It would also explain the absence of a \textbackslash (k=0\textbackslash )
rise in the methanol-octanol and octanol-octanol structure factors,
while methanol-methanol retains one: the octanol-rich organization
would remain comparatively stable, while the methanol-rich regions
fluctuate within it.

\section{Domain order characteristics\protect\label{sec:Domain-order-characteristics}}

In this section we examine the long-range domain correlations through
the running Kirkwood-Buff integrals (RKBI), defined as \cite{Perera2022}
\begin{equation}
G_{a_{i}b_{j}}(r)=2\pi\int_{0}^{r}sds[g_{a_{i}b_{j}}(s)-1]\label{RKBI}
\end{equation}

The RKBI amplify the weak long-range correlations in the tails of
$g(r)$, while their large-$r$ asymptotes provide the KBI values.
Their determination is particularly delicate in strongly structured
mixtures, where small deviations of $g(r)$ from its asymptote of
1 strongly affect the RKBI \cite{Perera2017,Perera2022,kbi-Nico,KBI-simonKurger,kbi-Dednam}.
We previously proposed a simple correction of this asymptote\cite{Perera2022},
which will be illustrated below where appropriate.

\subsection{Domain correlations and Kirkwood-Buff integrals\protect\label{subsec:Domain-correlations-and}}

\begin{figure}[H]
\centering
\includegraphics[scale=0.3]{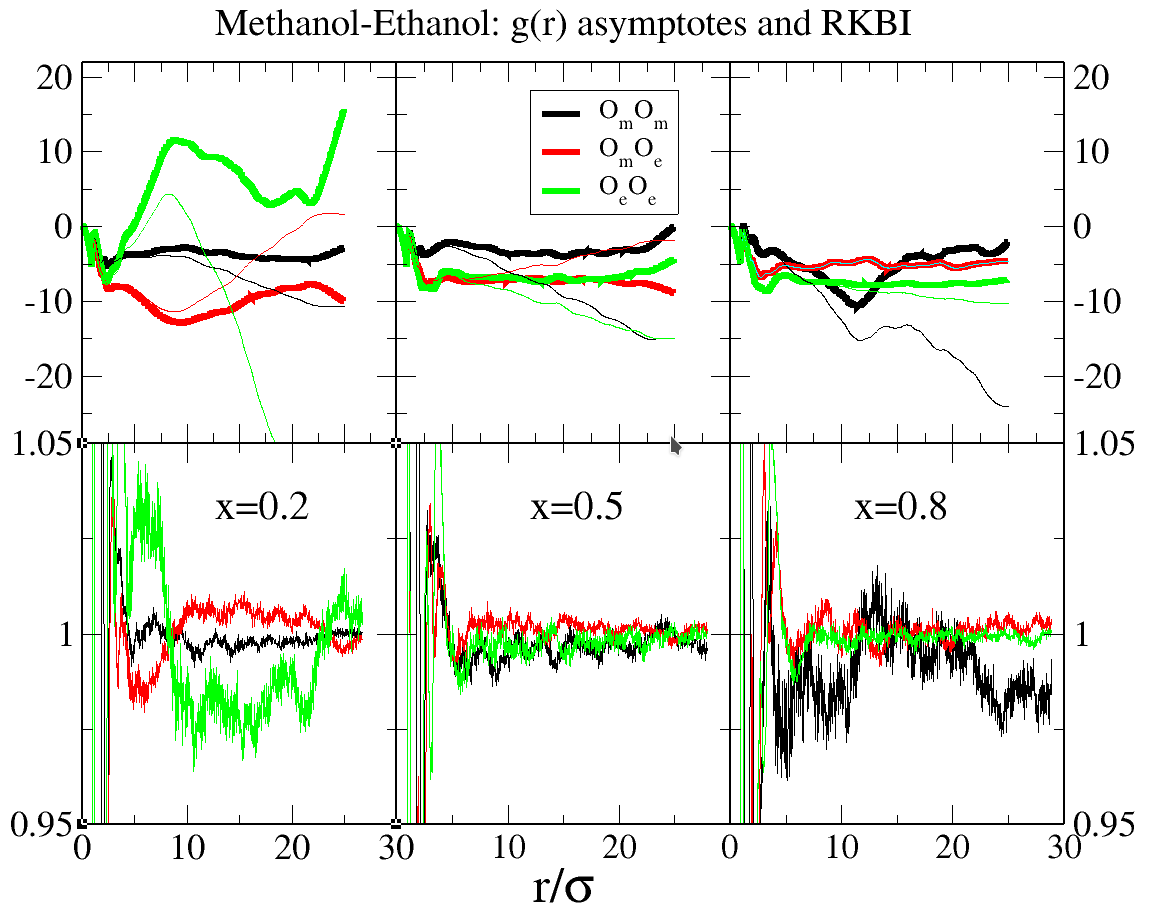}

\caption{Long range behaviour in methanol-ethanol oxygen-oxygen correlation
functions $g_{O_{a}O_{b}}(r)$ (lower row panels) and corresponding
RKBI (upper row panels, see text) and at typical ethanol mole fractions
$x=0.2$ (left panels), $0.5$(central panels) and $0.8$(right panels).
The color legends are given in the inset of the upper central panel
(black for methanol-methanol $g_{O_{m}O_{m}}(r)$, red for methanol-ethanol
$g_{O_{m}O_{e}}(r)$ and green for ethanol-ethanol $g_{O_{e}O_{e}}(r)$.}

\label{ME-KBI}
\end{figure}

The methanol-ethanol correlations are first reexamined at long range
in Fig. \ref{ME-KBI}. The lower panels magnify the weak tail oscillations
of $g_{O_{a}O_{b}}(r)$ around 1, while the corresponding RKBI in
the upper panels reveal their irregular character more clearly. We
observe an important feature, see also in the other mixtures studied
below: the tails of the g(r) have two level of noise. The first concerns
the smallest oscillations, which are naturally arising from statistics.
The second corresponds to a higher level oscillations which can be
linked to domain statistics. If domains were regular features, such
oscillations would have a single period, as observed for instance
in Ref.\cite{aup-POF}. This is clearly not the case here, hence refered
as the conundrum in Section \ref{subsec:The-conundrom-of}.

The thin curves in the upper panels show the uncorrected RKBI, whose
upward or downward drifts result from small deviations of $g(r)$
from its proper asymptote. 

Shifts of typically less than a few percent largely remove these drifts,
following the procedure documented in Refs. \cite{Perera2022,Lovrincevic2019}.
Yet even after this correction, the RKBI retain irregular oscillations
and no clear asymptotic value can be identified.

\begin{figure}[H]
\centering
\includegraphics[scale=0.3]{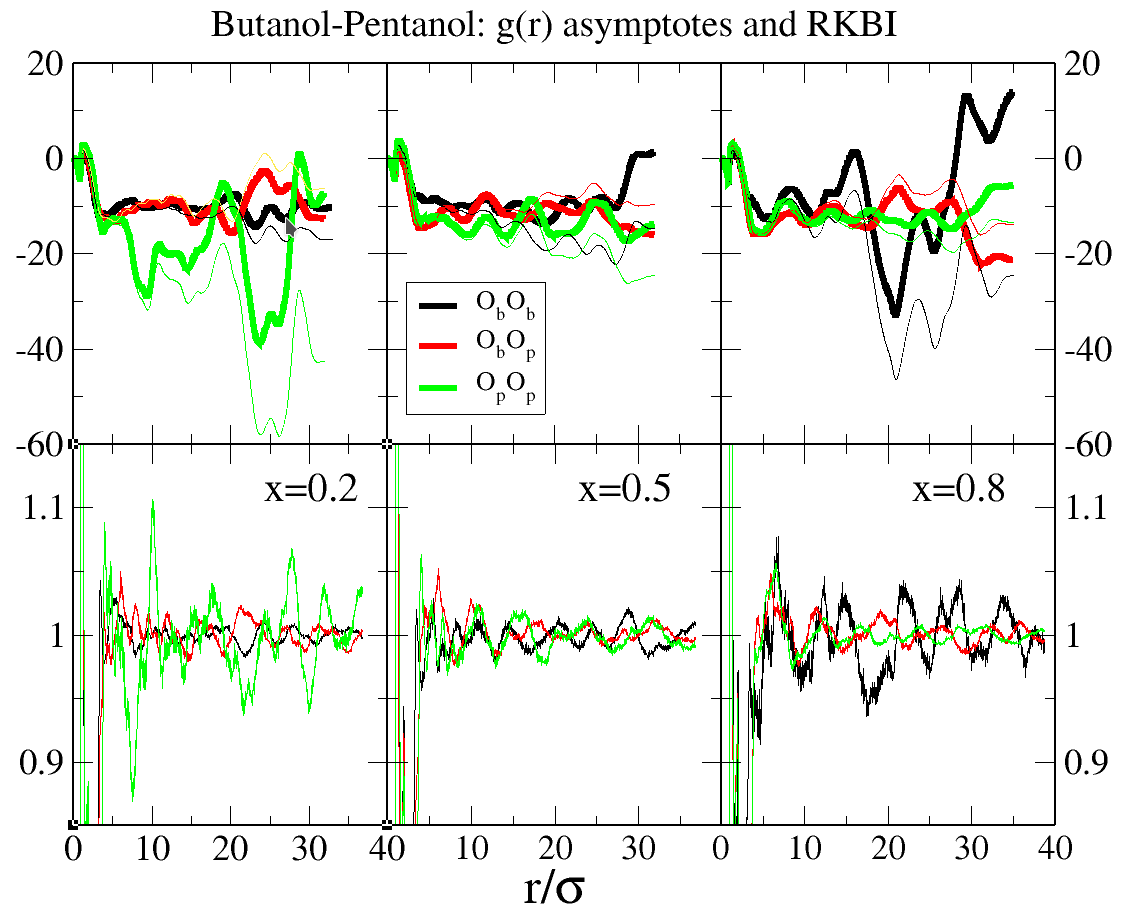}

\caption{Long range behaviour in butanol-pentanol oxygen-oxygen correlation
functions. Plotting conventions as in Fig.\ref{ME-KBI}.}

\label{BP-KBI}
\end{figure}

\begin{figure}[H]
\centering
\includegraphics[scale=0.3]{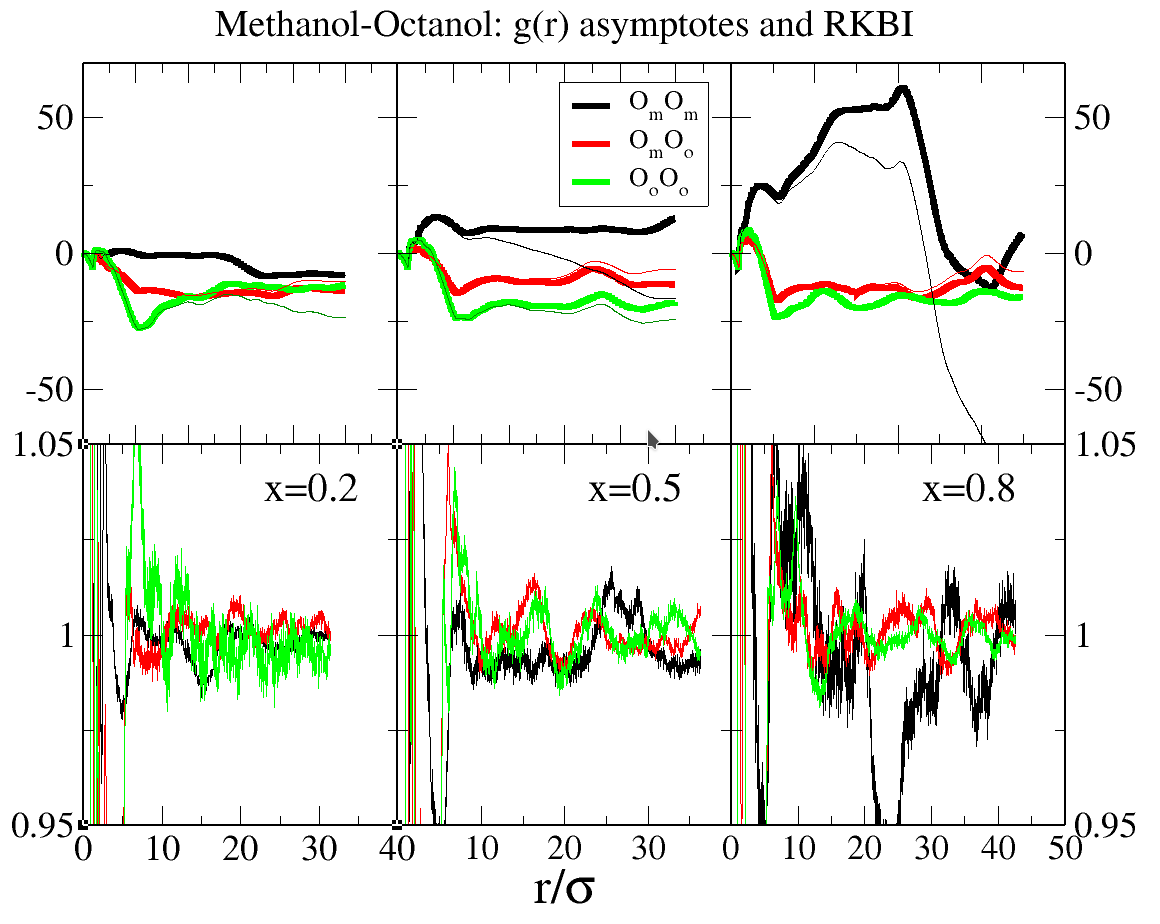}

\caption{Long range behaviour in methanol-octanol oxygen-oxygen correlation
functions. Plotting conventions as in Fig.\ref{ME-KBI}.}

\label{MO-KBI}
\end{figure}

The butanol-pentanol RKBI (Fig.\ref{BP-KBI}) display even more irregular
oscillations. In contrast, the methanol-octanol results (Fig. \ref{MO-KBI})
show better stabilized asymptotes for $x=0.2$ and $x=0.5$, from
which approximate KBI values can be extracted. 

Overall, however, the three systems display significant run-to-run
uncertainty in their long-range correlations, raising the question
of whether these oscillations eventually converge.

In these 3 figures, one notices the presence of significant noise
and uncertainty, and one may wonder how successive runs may affect
these artifacts. This is what we examine in the next part.

\subsection{Illustration of the conundrum in domain correlations}

\begin{figure}[H]
\centering
\includegraphics[scale=0.3]{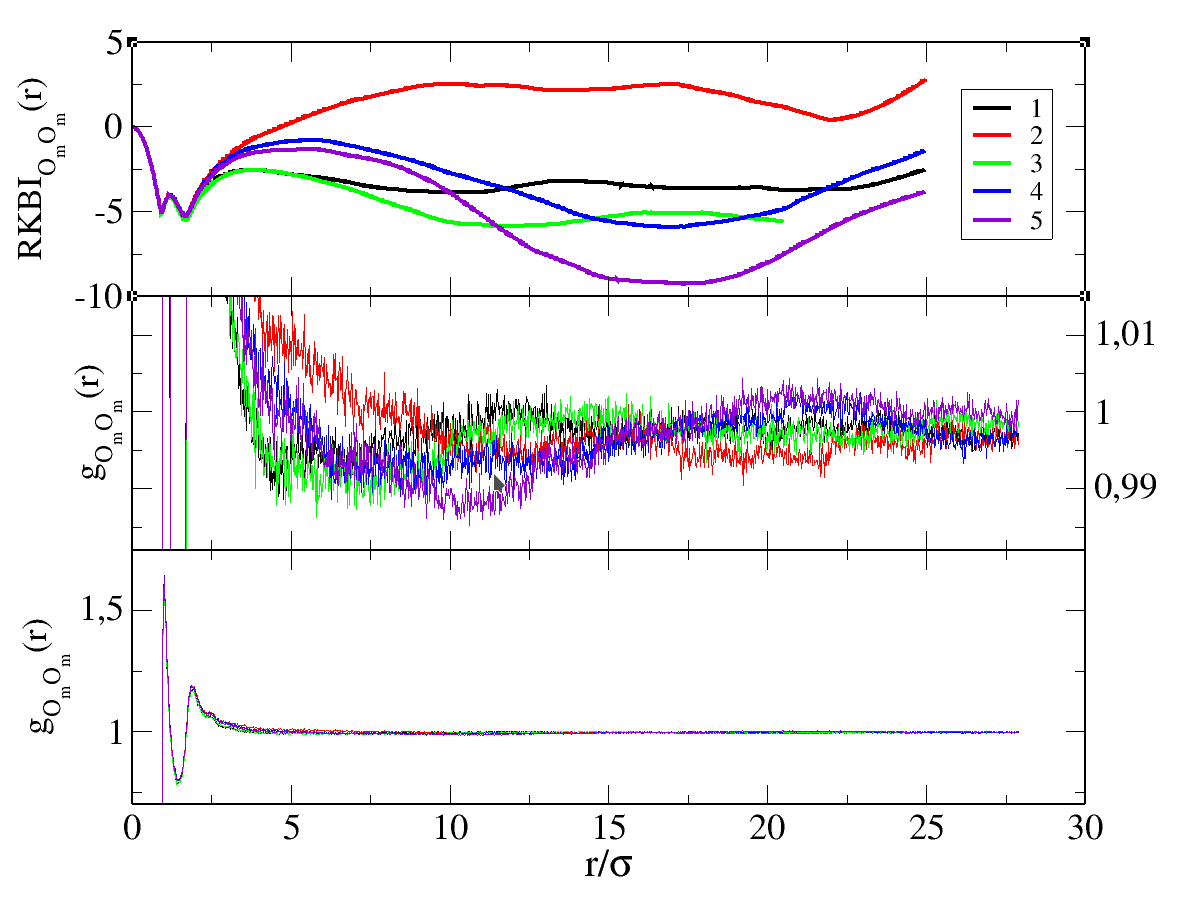}

\caption{Illustration of the lack of statistical convergence in the $g_{AB}(r)$tail
domain correlation for methanol-ethanol equimolar mixtures (x=0.5),
and shown here for the correlation function $g_{O_{m}O_{m}}(r)$ for
5 different successive runs, each for 150 thousand Monte Carlo moves
(each move concerns trial moves of all 1000 particles). The lower
panel shows all $g_{O_{m}O_{m}}(r)$ (unshifted- see text), the middle
panel is a zoom on the tail part showing domain oscillations, and
the upper panel shows the domain oscillations on the RKBI functions
(shifted - see text).}

\label{OmOm-KBI}
\end{figure}

Fig.\ref{OmOm-KBI} examines this question for the methanol-ethanol
mixture over five successive runs of 150 thousand Monte Carlo steps.
At ordinary scale, the $g_{O_{a}O_{b}}(r)$ are nearly indistinguishable.
Magnifying their tails, however, reveals both statistical noise and
the absence of a clearly converging oscillatory pattern. This becomes
even more apparent in the RKBI, which retain a global pattern but
do not settle towards a unique asymptotic behaviour.

This lack of convergence may simply reflect a numerical limitation,
potentially removable with larger systems or longer sampling. We propose,
however, that it may also reflect an intrinsic property of micro-heterogeneous
systems. If each domain is regarded as a loose \textquotedblleft particle\textquotedblright{}
of ill-defined size, the long-range oscillations represent correlations
between such domain-particles. A mean domain size may still be defined,
for example from a pre-peak position, while correlations between individual
domains remain affected by their intrinsically ill-defined particle
character.

This viewpoint may be relevant more generally to soft-matter systems,
where several levels of supramolecular structure coexist and their
correlations may raise similar statistical problems in computer simulations.

\section{Discussion }

The present study spans several levels of description, from the physical
chemistry of associating liquids to more fundamental questions of
statistical physics. Although simplified, the 2D models retain characteristic
features of their real 3D counterparts, such as the near-ideal mixing
of similar alcohols, while providing easier access to long-range fluctuations
and structural correlations.

Concerning the irregular domain-like oscillations observed in the
long range part of the g(r), these were initially observed in smaller
systems of N=500 particules. Clearly, going to N=1000 did not change
these features, which are equally observed for the few N=2000 particles
we studied. We conclude that only mesoscopic size simulations could
achieve proper sampling. Although it would be tempting to ``pay the
price'' of such simulations, we propose that the difficulties in
propely sampling the KBI are an intrinsic property of systems that
develop fluctuations that present a duality between concentration
fluctuations and micro-heterogeneity\cite{Pozar2016a,myIUPAC,aup-POF}.

The importance of fluctuations in 2D brings these mixtures close to
other systems studied in statistical physics, such as unoriented nematics
\cite{Gennes1993} or the 2D XY model \cite{chaikin2000principles}.
In the former, the director may meander throughout the sample, making
a mean orientation difficult to define; similarly, the fluctuating
domains observed here resist a simple mean description. In the XY
model, topological defects emerge as effective Coulomb charges, although
each defect is itself embedded in an extended orientational structure.
These analogies suggest that the difficulties encountered here may
benefit from field-theoretical descriptions. In recent work \cite{aup-POF},
we showed how microscopic liquid-state theory can be connected to
field theory through the mesoscopic contribution to the bridge function.

\section{Conclusion}

We have investigated binary mixtures of two-dimensional site-based
alcohol models, focusing on the interplay between mixing, chain-like
aggregation, and spatial correlations. The reduced dimensionality
preserves essential features of associating liquids, such as charge
ordering of the hydroxyl heads, while allowing structural organization
to be followed over a broad range of length scales.

A central result is that mixtures which phase-separate in three dimensions
remain macroscopically mixed in two dimensions over the compositions
investigated. This mixing nevertheless retains a strong local organization,
dominated by chain-like aggregates of the polar heads. At larger scales,
the snapshots reveal micro-segregated patterns, while the structure
factors show either a $k=0$ rise or no separate domain pre-peak.
In some cases, the characteristic domain and aggregate length scales
may therefore overlap, making them difficult to distinguish in reciprocal
space.

The weak long-range correlations also have direct consequences for
the running Kirkwood-Buff integrals, whose asymptotic behaviour is
difficult to establish within the accessible system sizes and sampling
times. While larger systems or longer simulations may ultimately recover
well-defined KBI values, we propose that part of this difficulty may
be intrinsic to domains behaving as loose objects of fluctuating size
and connectivity.

More generally, these results emphasize that macroscopic miscibility,
local aggregation, and concentration fluctuations need not correspond
to distinct structural regimes. In micro-heterogeneous liquids, they
may coexist over different, and sometimes overlapping, length scales.
This makes simple fluctuation-based descriptions difficult to apply
and points towards descriptions in which the hierarchical organization
of transient aggregates is explicitly taken into account.

\section*{Supporting Information}

The Supporting Information document contains various figures complementary
to those of the main paper, as well as tables of the force field charge
parameters.

\bibliographystyle{jpcb_final.bst}
\bibliography{alc2Dmix}

\end{document}